\documentclass[sigconf]{acmart}
\usepackage{textcomp}
\usepackage{enumitem}
\usepackage{color,soul}
\usepackage[font=scriptsize]{caption}
\usepackage[compact]{titlesec}
\titlespacing{\subsubsection}{0pt}{1.5ex}{1ex}
\usepackage{bm}
\usepackage{multirow}
\usepackage{multicol}
\usepackage{adjustbox}
\usepackage{tabularx}
\usepackage{etoolbox}
\usepackage{subcaption}
\usepackage{multirow}
\usepackage{multicol}
\usepackage{graphicx}
\usepackage{titlesec}
\usepackage{pbox}
\usepackage{caption}
\usepackage{appendix}
\usepackage{tikz}

\AtBeginDocument{%
  \providecommand\BibTeX{{%
    \normalfont B\kern-0.5em{\scshape i\kern-0.25em b}\kern-0.8em\TeX}}}

\setcopyright{rightsretained}
\acmISBN{}
\acmDOI{}
\acmYear{}
\acmConference[Marble-KDD '21]{Marble-KDD '21}{August 16, 2021}{Singapore}
\begin{document}

%%
%% The "title" command has an optional parameter,
%% allowing the author to define a "short title" to be used in page headers.
\title{High Quality Related Search Query Suggestions using Deep Reinforcement Learning}

%%
%% The "author" command and its associated commands are used to define
%% the authors and their affiliations.
%% Of note is the shared affiliation of the first two authors, and the
%% "authornote" and "authornotemark" commands
%% used to denote shared contribution to the research.
\author{Praveen Kumar Bodigutla}
%\authornote{Both authors contributed equally to this research.}
%\email{anonymous@anonymous.com}
%\orcid{1234-5678-9012}
%\author{******}
%\authornotemark[1]
\email{pbodigutla@linkedin.com}
\affiliation{%
  \institution{LinkedIn}
  \streetaddress{1000 W Maude Ave}
  \city{Sunnyvale}
\state{California}
\country{USA}
  \postcode{94085}
}

\begin{abstract}
``High Quality Related Search Query Suggestions''  task aims at recommending search queries which are real, accurate, diverse, relevant and engaging. Obtaining large amounts of query-quality human annotations is expensive. Prior work on supervised query suggestion models suffered from selection and exposure bias, and relied on sparse and noisy immediate user-feedback (e.g., clicks), leading to low quality suggestions. Reinforcement Learning techniques employed to reformulate a query using terms from search results, have limited scalability to large-scale industry applications.

To recommend high quality related search queries, we train a Deep Reinforcement Learning model to predict the query a user would enter next. The reward signal is composed of long-term session-based user feedback, syntactic relatedness and estimated naturalness of generated query. Over the baseline supervised model, our proposed approach achieves a significant relative improvement in terms of recommendation diversity ($3\%$), down-stream user-engagement ($4.2\%$) and  per-sentence word repetitions ($82\%$).

\end{abstract}

%%
%% The code below is generated by the tool at http://dl.acm.org/ccs.cfm.
%% Please copy and paste the code instead of the example below.
%%

%\ccsdesc[300]{Computing methodologies~Natural language generation}

%%
%% Keywords. The author(s) should pick words that accurately describe
%% the work being presented. Separate the keywords with commas.

%% A "teaser" image appears between the author and affiliation
%% information and the body of the document, and typically spans the
%% page.
%\begin{teaserfigure}
%  \includegraphics[width=\textwidth]{sampleteaser}
%  \caption{Seattle Mariners at Spring Training, 2010.}
%  \Description{Enjoying the baseball game from the third-base
%  seats. Ichiro Suzuki preparing to bat.}
%  \label{fig:teaser}
%\end{teaserfigure}

%%
%% This command processes the author and affiliation and title
%% information and builds the first part of the formatted document.
\maketitle

\section{Introduction}
Related search query suggestions aims at suggesting queries that are related to the user's most recent search query. For example, search query ``machine learning jobs'' is related to ``machine learning''. These suggestions are important to improve the usability of search engines \cite{10.1145/1401890.1401995}. We define high-quality related search query suggestions as query recommendations which are natural (i.e., entered by a real user), diverse, relevant, error-free and engaging.

Sequence-to-Sequence (Seq2Seq) encoder-decoder architectures are widely used to generate query suggestions \cite{wang2020deep,nogueira-cho-2017-task,Yu2017SeqGANSG,DBLP:conf/cikm/KaziGGL20}. Supervised autoregressive generative models trained with ground-truth labels suffer from exposure bias \cite{Yu2017SeqGANSG, schmidt-2019-generalization}. Maximum Likelihood Estimation (MLE) training generates repetitive sequences \cite{DBLP:journals/corr/LiMRGGJ16}.  Machine learning approaches \cite{DBLP:conf/cikm/KaziGGL20,10.1145/3308558.3313412} that are trained on immediate user feedback, such as click on recommended query, are prone to selection-bias \cite{45286}. Reinforcement Learning techniques were proposed to address these limitations \cite{wang2020deep}.  

``Local'' Deep-Reinforcement-Learning (DRL) frameworks for query reformulation \cite{wang2020deep,nogueira-cho-2017-task} adjust a query relative to the initial search results. Reward signals used in these approaches were computed from mined search result documents.  Processing search results to reformulate the query and mining entire collection of documents is not practical in large-scale real-world applications, due to rapidly changing and ever-growing user-generated content. Hence, we employ ``global'' DRL approaches \cite{nogueira-cho-2017-task}, that are independent of the set of documents returned by the original query and instead depend on the actual queries entered by the users and the feedback provided by them. Furthermore, we choose the more useful approach to predict the queries a user will enter next, over simply reformulating the current query \cite{10.1145/1401890.1401995}. We mine search sessions and corresponding co-occuring query pairs from LinkedIn search engine logs. Unlike general purpose web search, our work is focused on domain-specific-search queries \cite{cite-key-dss}, that users enter to search for job postings, people profiles, user-groups, company pages and industry news. 

%Broadly, prior methods to suggest related queries focused on two approaches: 1) reformulate the original query \cite{wang2020deep,ponnusamy2019feedbackbased,nogueira-cho-2017-task}; or 2) suggest queries that the user will enter next \cite{10.1145/1401890.1401995,DBLP:conf/cikm/KaziGGL20}. The latter approach is more useful than simpily replacing the current query \cite{10.1145/1401890.1401995}. In order to solve the multi-objective high-quality related search query suggestions problem, our work focuses on generating queries which the user will enter next.

DRL text generation approaches such as policy-gradient based Sequential Generative Adversarial Networks (SeqGAN) \cite{Yu2017SeqGANSG} achieved excellent performance on generating creative text sequences. Most recently, on summary generation task, fine-tuning pre-trained supervised model using Proximal Policy Optimization (PPO) \cite{DBLP:journals/corr/SchulmanWDRK17} method outperformed supervised GPT3 \cite{brown2020language} model 10x its size \cite{DBLP:journals/corr/abs-2009-01325}.  Motivated by the performance improvements achieved by DRL techniques on text-generation problems, we solve the high quality query suggestions problem by modeling the query generator as a stochastic parametrized policy. Specifically, we fine-tune the state-of-the-practice Seq2Seq Neural Machine Translation (NMT) model \cite{Bahdanau2015NeuralMT} for query generation \cite{DBLP:conf/cikm/KaziGGL20}, using policy-gradient REINFORCE \cite{cite-key-reinforce} algorithm. Seq2Seq NMT models are widely popular in industry applications, especially in low resource environments\cite{tushan2020}. 

Supervised reward estimation models are commonly used to optimize text-generation policy \cite{wang2020deep,nogueira-cho-2017-task,DBLP:journals/corr/abs-2009-01325,Yu2017SeqGANSG}. SeqGAN model used GAN  \cite{10.5555/2969033.2969125} discriminator output as reward signal. The supervised discriminator model predicted if the generated sequence is ``real''. The estimated reward is passed back to the intermediate state-action steps using computationally expensive Monte Carlo search \cite{10.5555/3022539.3022579}. PPO based summary generation model  \cite{DBLP:journals/corr/abs-2009-01325} used estimated annotated user feedback on generated summaries as the reward signal. Immediate implicit user feedback is sparse, asking users to provide explicit feedback is intrusive \cite{bodigutla2019domainindependent} and reliable annotations are expensive to obtain. 

In our proposed approach, the DRL future-reward is composed of three signals, which are: 1) long-term implicit user-feedback within a search-session; 2) unnatural query generation penalty; and 3) syntactic similarity between generated query and user's most recent search query. Leveraging implicit user-feedback from search-sessions as opposed to using immediate feedback, helps in maximizing user engagement across search-sessions, addresses the reward-sparsity problem and removes the need to obtain expensive human annotations (Section \ref{sec:reward-formulation}). We design a weakly supervised context-aware-naturalness estimator model, which estimates the naturalness probability of a generated query. Similar to \cite{Yu2017SeqGANSG} we perform Monte Carlo search to propagate rewards. However, we reduce the computation cost considerably by performing policy roll-out only from the first time-step of the decoder (Section \ref{sec:drl-model}). 

In summary, we employ DRL policy-gradient technique for making high-quality related-query suggestions at scale. Our proposed approach achieves improvement in-terms of recommendation diversity, down-stream user-engagement, relevance and errors per sentence. To the best of our knowledge, this is the first time a combination of long-term session-based user-feedback, un-natural sentence penalty and syntactic relatedness reward signals are jointly optimized to improve query suggestions' quality.

The remainder of this paper is structured as follows: Section \ref{sec:approach} describes our proposed deep reinforcement learning approach. Section \ref{sec:experiment-setup} presents the experimental setup and discusses empirical results. Section \ref{sec:conclusion} concludes.

\section{Approach for Improving Query Suggestions}
\label{sec:approach}
We fine-tune a weakly supervised Sequence-to-Sequence (Seq2Seq) Neural Machine Translation (NMT) model ($Seq2Seq_{NMT}$) to initialize the query generation policy. The process then consists of two steps: 1) Learn a context-aware weakly supervised naturalness estimator;  and 2) Fine-tune pre-trained supervised $Seq2Seq_{NMT}$ model using REINFORCE \cite{cite-key-reinforce} algorithm. The future-reward is composed of user-feedback in a search session ($U_{+}$), syntactic similarity ($ROUGE$) and unnaturalness penalty ($- \eta * (1-D_\phi)$) of the generated query given the co-occuring previous query.

%\vspace{-0.2cm}
\begin{figure}[h]
\includegraphics[width=0.45\textwidth]{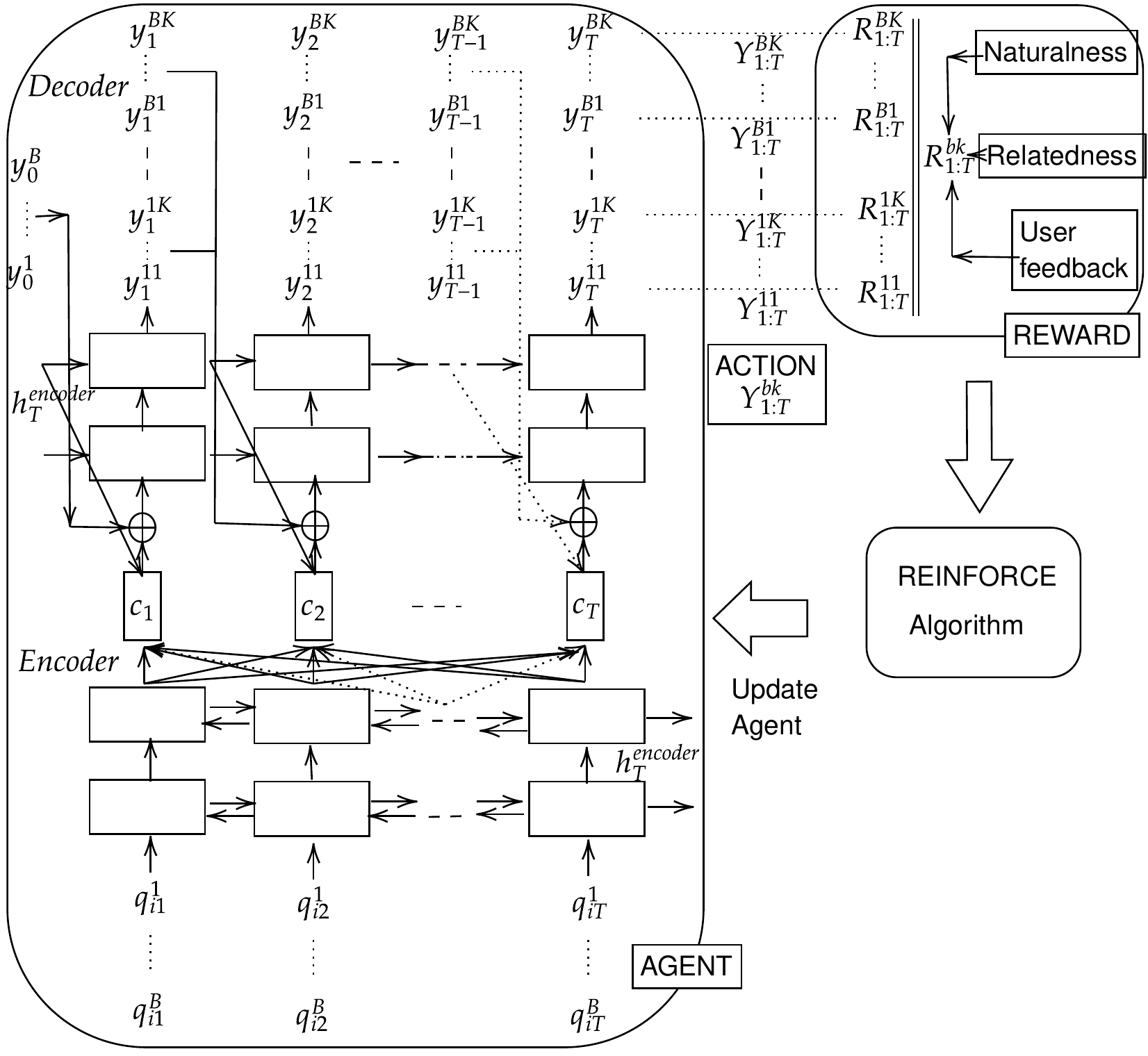}
%\vspace{-0.4cm}
\caption{
\textmd{Deep Reinforcement Learning for text-generation using REINFORCE algorithm. Agent is an encoder-decoder Seq2Seq attention model. $q^{bk}_{i\{1:T\}}$ is the input query, for batch-index $b \epsilon B$ and Monte-Carlo sample index $k \epsilon K$. Generated words ($y^{{1:B},1:K}_t$) concatenated with attention context ($c_{t+1}$) is passed as input to the decoder's $t+1$ time-step. For each generated sequence ($y^{bk}_{1:T}$) policy update is done using RL reward ($R^{bk}_{1:T}$), which is calculated at the end of each generated sequence.}}
\label{fig:model}
\end{figure}
%\vspace{-0.4cm}

\subsection{Weakly Supervised Pre-Training}
\label{sec:supervised}
Variants of mono-lingual supervised $Seq2Seq_{NMT}$ models are used in industry applications for query suggestions [14]. In the pre-training step, we train the supervised $Seq2Seq_{NMT}$ model using co-occurring consecutive query pairs in a search sessions. A search-session \cite{10.1145/1401890.1401995} is a stream of queries entered by a user in a 5-min\footnote{Based on guidance from internal search team's proprietary analysis.} time window. N-1 Consecutive query pairs ($q_i$, $q_{i+1}$) are extracted from a search session consisting of a sequence of N queries ($q_1$, $q_2$,...,$q_N$). Consecutive queries could be unrelated, semantically and (or) syntactically related. Our model is weakly supervised as we use all query pairs and do not filter them using sparse click data, costly human-evaluations and weak association rules. Weak supervision allows the training process to scale, minimize selection-bias and we conjecture that it improves model generalization too. For example, unlike [14], we do not apply syntactic similarity heuristics to filter query pairs, as queries could be semantically related yet syntactically dissimilar (e.g., ``artificial intelligence'' and ``machine learning'') . The $Seq2Seq_{NMT}$ encoder-decoder framework consists of a BiLSTM \cite{6707742} encoder, that encodes a batch (batch-size $B$) of input queries ($q^{1:B}_i$) and the LSTM \cite{10.1162/neco.1997.9.8.1735} decoder generates a batched sequence of words y = ($y^{1:B}_1$,..., $y^{1:B}_T$). Where, $T$ is the sequence length. During training, we use teacher forcing \cite{Williams89alearning}, i.e., use the co-occuring query ($q^{1:B}_{i+1}$) as input to the decoder. Context attention vector is obtained from the alignment model \cite{Bahdanau2015NeuralMT} . Categorical cross entropy loss is minimized during training and hyper-parameters of the model are fine-tuned (see Section \ref{sec:experiment-setup}).

\subsection{Fine-tuning using Deep Reinforcement Learning}
\label{sec:drl}
This section describes the reward estimation and Deep Reinforcement Learning (DRL) model training steps to fine-tune and improve the policy obtained via pre-trained supervised model.

\subsubsection{Deep Reinforcement Learning Model.} 
\label{sec:drl-model}
 \hspace{0.2cm}Parameters of the DRL agent $G_{\theta}$ are initialized with pre-trained $Seq2Seq_{NMT}$ model (Section \ref{sec:supervised}). The initial policy is fine-tuned using the REINFORCE policy-gradient algorithm (Figure \ref{fig:model}). `$K$' complete sentences ($y^{b,{1:K}}_{1:T}$) generated per query ($q^{b}_i$) constitute the action space at time-step $T$, where $b \epsilon B$ is the index in a mini-batch of $B$ queries. To mitigate exposure-bias, generated words ($y^{1:B, 1:K}_{t-1}$) from previous time-step are passed as input to the next time-step `$t$' of the decoder. Future-reward ($R_{D_\phi}(y^{bk}_{1:T})$) computed at the end of each generated sample, is back-propagated to the encoder-decoder model. Given the start-state ($S^{b}_0$) comprising of the input query ($q_i^b$) and <START> token $y^{b}_{0}$, the objective of the agent is to generate related-search query suggestions ($y^{b}_{1:T}$) which maximize objective: 
\setlength\abovedisplayskip{0pt}
\begingroup
 \begin{equation}
 \label{eq:objective}
 J(\theta) = \mathbb{E}[R_{D_\phi}(y^{b}_{1:T})|S^{b}_0, \theta]\\\
\scriptsize
%\small
 \end{equation}
Where per-sample reward is:
\begin{equation}
\label{eq:reward}
R_{D_\phi}(y^{bk}_{1:T}) = U_{+}  + (1- U_{+})*(ROUGE_{q^b_i, y^{bk}_{1:T}} - \eta*(1-D_\phi(y^{bk}_{1:T}))) 
\scriptsize
%\small
\end{equation}
\endgroup
MC approximation of the gradient using likelihood ratio trick:
\begin{equation}
\label{eq:gradient}
\Delta_{\theta} J(\theta) \approx \frac{1}{K} * \Sigma_{k \epsilon K} [R_{D_\phi}(y^{bk}_{1:T}) * \Delta_{\theta} log\,G_{\theta}(y^{bk}_{1:T}|S^{b}_0)], \,\,\, y^{bk}_{1:T} \,\epsilon\, MC^{G_{\beta}}(y^{bk}_{1:T})
\scriptsize
%\small
\end{equation}
%Action-Value function is defined as:
%\begin{equation}
%\label{eq:q-reward}
%\Delta_{\theta} J(\theta) = \mathbb{E}_{(y^{b}_{1:T}  \textasciitilde{G_{\theta}})} [Q^{G_{\theta}}_{D_\phi} (y^{b}_{1:T}) * \Delta_{\theta} G_{\theta}(y^{b}_{1:T}|S^{b}_0) ]
%Q^{G_{\theta}}_{D_\phi} (y^{b}_{1:T}) = 1/K * \Sigma_{k \epsilon K} R_{D_\phi}(y^{bk}_{1:T}), \,\,\, y^{bk}_{1:T} \,\epsilon\, MC^{G_{\beta}}(y^{bk}_{1:T})
%\small
%\end{equation}
Unlike SeqGAN \cite{Yu2017SeqGANSG} DRL model training, where the action-value at each intermediate time-step $t$ is evaluated by generating $K$ samples ($y^{b,1:K}_{t+1:T}$), we perform MC policy roll-out from the start state alone. $K$ queries ($y^{b,1:K}_{1:T}$) are generated using roll-out policy $G_\beta$, for each input query $q_i^b$. This modification reduces the computation cost by a factor of $\mathcal{O}(T)$  ($\mathcal{O}(KT^2) \rightarrow \mathcal{O}(KT)$) per input query. $G_\beta$ is initialized with $G_\theta$ and is periodically updated during training, using a configurable schedule (see Algorithm 1). At each time-step $t$, the state $S_t^{bk}$ is comprised of the input query and the tokens produced so far ($\{q_i^b, y^{bk}_{1:t-1}\}$) and the action is the next token $y^{bk}_t$ to be selected from stochastic policy $G_\theta(y^{bk}_t | S_t^{bk})$. 

Details of the constituents of future-reward ($U_{+}$, $ROUGE_{q^b_i, y^{bk}_{1:T}}$ and $D_\phi(y^{bk}_{1:T})$) are in the next section (Section \ref{sec:reward-formulation}).  Since the expectation $\mathbb{E}[.]$ can be approximated by sampling methods, we then update the generator’s parameters with $\alpha$ as the learning-rate as: 

%\vspace{-0.2cm}
\begingroup
\begin{equation}
\label{eq:sgd}
\theta  \leftarrow \theta + \alpha * \Delta_{\theta} J(\theta) 
\scriptsize
% \small
\end{equation}

\resizebox{0.43\textheight}{!}{
%\vspace{-0.2cm}
\begin{adjustbox}{max width=\textwidth}
\bgroup
\tikzset{every picture/.style={line width=0.48pt}} %set default line width to 0.75pt        
\label{fig:algorithm}
\begin{tikzpicture}[x=0.75pt,y=0.75pt,yscale=-1,xscale=1]
%uncomment if require: \path (0,513); %set diagram left start at 0, and has height of 513
%Straight Lines [id:da7069235011335131] 
\draw    (23,13) -- (465,13) ;
%Straight Lines [id:da12281057385598837] 
\draw    (23, 390) -- (465, 390) ;
% Text Node
\draw (24,23) node [anchor=north west][inner sep=0.75pt]   [align=left] {\textbf{Require:} Generator policy $\displaystyle G_{\theta }$ };
% Text Node
\draw (175,23) node [anchor=north west][inner sep=0.75pt]   [align=left] {; roll-out policy $\displaystyle G_{\beta }$ \ \ };
% Text Node
\draw (270,24) node [anchor=north west][inner sep=0.75pt]   [align=left] {;};
% Text Node
\draw (280,24) node [anchor=north west][inner sep=0.75pt]   [align=left] {naturalness-estimator $\displaystyle D_{\phi }$};
% Text Node
\draw (405,26) node [anchor=north west][inner sep=0.75pt]   [align=left] {;};
% Text Node
\draw (25,43) node [anchor=north west][inner sep=0.75pt]   [align=left] {Query-pair in a search session $\displaystyle \left( q_{i\{1:T\}}^{1:B} ,\ q_{\{i+1\}\{1:T\}}^{1:B}\right)$; Batch size: B; };
% Text Node
\draw (24,67) node [anchor=north west][inner sep=0.75pt]   [align=left] {MC sampling-strategy $\displaystyle \epsilon \ $[beam-search, sampling from categorical distribution]};

% Text Node
\draw (315,260.44) node   [align=left] {\begin{minipage}[lt]{430.89pt}\setlength\topsep{0pt}
1. \quad Fine-tune supervised model using MLE with$\displaystyle \ q_{i\{1:T\}}^{1:B} \ $as input sequence and \\  \null \quad  \quad $\displaystyle  q_{\{i+1\}\{1:T\}}^{1:B} \ $as target sequence.\\2. 	\quad Initialize $\displaystyle G_{\theta }$ with fine-tuned supervised model.\\3. \quad	$\displaystyle \beta \ \leftarrow \ \theta $	\\4.\quad	Train contextual-naturalness-estimator $\displaystyle D_{\phi} $ using negative examples generated from $\displaystyle  G_{\theta }$\\5. \quad	\textbf{repeat}\\6.	\quad\quad	\textbf{for} n steps\\7.			\quad \quad \quad \textbf{foreach} $\displaystyle b\ \epsilon \ B$\\8.	\quad \quad \quad \quad	Generate "K" sequences $\displaystyle y_{1:T}^{b,\ 1:K}$ using configured sampling-strategy \\		\null\quad \quad \quad \quad \quad			$\displaystyle y_{1:T}^{b,\ 1:K}$ = $\displaystyle \left( y_{1}^{b,\ 1:K} ,\ y_{2}^{b,\ 1:K} ,\ ...,y_{T}^{b,\ 1:K}\right) \ \sim \ G_{\beta } \ \ \ $\\9.	\quad \quad \quad \quad		Compute future-reward at the end of each generated sequence\\ 		\null\quad \quad \quad \quad \quad			$\displaystyle R_{1:T}^{b,k} =\ U_{+} \ +\ ( 1\ -\ U_{+}) \ *\ ( ROUGE( q_{i\{1:T\}}^{b}$,$\displaystyle y_{\{1:T\}}^{b,k})  \ -\ \eta *( 1-D_{\phi }))$\\ 10. \quad \quad \quad \hspace{0.1cm} Compute gradient $\Delta J(\theta)$ (Equation \ref{eq:gradient}) \\11. \quad \quad \quad \hspace{0.1cm} Update $\displaystyle G_{\theta }$ parameters via policy-gradient (Equation 4) \\12.		\quad \quad \quad 	\textbf{end foreach}\\13.	\quad \quad	\textbf{end} for\\14. 		$\displaystyle  \quad \quad	\beta \ \leftarrow \ \theta $\\15. \quad	\textbf{until} convergence\\\\\\ \ \ \ \textbf{Algorithm 1:} Deep Reinforcement Learning algorithm for related-search query suggestions. \\

\end{minipage}};

\end{tikzpicture}
\egroup
\end{adjustbox}
}

%\vspace{-0.3cm}
\subsubsection{Reward Formulation.} 
\label{sec:reward-formulation}
 \hspace{0.2cm}This section describes the three components of the future-reward $R_{D_\phi}(y^{bk}_{1:T})$ , which are session based positive user-feedback ($U_{+}$), syntactic-similarity ($ROUGE_{q^b_i, y^{bk}_{1:T}}$) and unnatural suggestion penalty ($- \eta*(1-D_\phi(y^{bk}_{1:T}))$).

\begin{itemize} [leftmargin=*]
\item \textbf{Long-term user-feedback in a search session ($U_{+}$)}: Viewing search-results (dwell time >= 5 seconds \footnote{Determined by internal domain-specific search user-behavior analysis.}) or performing an engaging action such as: sending a connection request or a message; applying for a job; favoriting a group, profile or an article, constitute positive implicit user actions in a search-session. Immediate user-feedback is sparse. However, positive user action percentage increases by absolute $11\%$\footnote{Computed over a set of \textasciitilde{100} million query pairs extracted from search-sessions in one month window. Actual $\%$ of positive user actions is not included due to confidentiality.}, when the remainder of the search session after the user enters $q_{i+1}$ is considered. In our work we maximize session-based user-feedback, as we are interested in maximizing user engagement across search sessions. For a generated query $y_{1:T}^{bk}$, session-based user feedback ($U_{+}$) is ``1'' , if a positive down-stream user action is observed in the remainder of the search-session and ``0'' otherwise.

\item \textbf{Relatedness of generated query to source query ($ROUGE_{q^b_i, y^{bk}_{1:T}}$)}: Despite increasing the percentage of associated positive user action by considering user's feedback across a search session, the label sparsity problem is not completely mitigated. In the search query logs, when there is no positive downstream user action associated with a generated query $y^{bk}_{1:T}$, we estimate the reward using a syntactic similarity measure. Reformulated queries are syntactically and semantically similar \cite{lee2021contrastive}. We compute syntactic relatedness of generated query ($y^{bk}_{1:T}$) with the source query ($q^b_i$) using ROUGE-1 \cite{lin-2004-rouge} score.

\item \textbf{Naturalness probability of generated query ($D_\phi(y^{bk}_{1:T})$)}: Users enter either natural language search queries\cite{Borges2020QueryUF} (e.g., \textit{``jobs requiring databases expertise in the bay area''} ) or they just enter key-words (e.g.,  \textit{``bay area jobs database''} ). In the context of related query suggestions, we define a ``natural''  query as one which a real user is likely to enter. We train a contextual-naturalness-estimation model (see Section \ref{sec:naturalness}) to predict naturalness probability $D_\phi(y^{bk}_{1:T})$ of a generated query, given the previous query entered by the user as context. \textit{``AI jobs''} is an example of a natural query after the user searched for \textit{``Google''}, even though both queries are syntactically and semantically (jobs vs company) dissimilar. However, \textit{``AI AI jobs jobs''} is unnatural and is unlikely to be entered by a real user. In our DRL reward formulation, we add penalty term $-\eta*(1-D_\phi(y^{bk}_{1:T}))$ to syntactic-relatedness ($ROUGE_{q^b_i, y^{bk}_{1:T}}$) score to discourage generation of unnatural queries. Coefficient $\eta$ is the configurable penalty weight. 
\end{itemize}

\subsubsection{Contextual Naturalness Estimator.}
\label{sec:naturalness}
\hspace{0.2cm}Our proposed contextual-naturalness-estimator is a BiLSTM \cite{6707742} supervised model, which predicts the probability a generated query is ``natural'' ($D_\phi(y^{bk}_{1:T})$) . Concatenated with query-context ($q_i$), user entered queries ($q_{i+1}$) serve as positive examples ($q_{i} \oplus q_{i+1}$) to train the model. We employ four methods to generate negative examples ($q_{i} \oplus q^{neg}_{i}$) per each positive example, which are: 1) With $q_i$ as input, sample query $q^{neg}_i$ from fine-tuned $Seq2Seq_{NMT}$ model's decoder (Section \ref{sec:supervised}); 2) perturb $q_{i+1}$ by duplicating a word in randomly selected position within the sentence; 3) replace a word with unknown word token (``<UNK>'') at a randomly selected position in $q_{i+1}$; and 4) generate a sentence by repeating a sampled word from a categorical distribution ($p_{w_1}...p_{w_{|V|}}$) for (randomly chosen) $r \epsilon [1, T-1]$ times. |V| is the size of the training data vocabulary, $p_{w_i}$ = $n_{w_i}/N_{|V|}$, $n_{w_i}$ is word ($w_i$) frequency and $N_{|V|} = \Sigma_{w_i\epsilon|V|} n_{w_i}$. Methods $1$, $2$ and $3$ generate hard-negative examples \cite{lee2021contrastive} and $4$ captures popularity-bias \cite{popularity-bias}, a situation where popular terms are generated more often than terms in long-tail.

To validate our hypothesis that sampled queries from $Seq2Seq_{NMT}$ are less natural than the ones provided by the user, we asked three annotators to rate $100$ randomly sampled query pairs ($q_1, q_2$). Query $q_1$ is entered by the user and $q_2$ is either sampled from a supervised model ($46\%$)  or was entered by the user ($54\%$) after searching for $q_1$. Without revealing the source of $q_2$, we asked annotators to identify if the query is ``natural'' (defined in Section \ref{sec:reward-formulation}) . On an average $58\%$ of model-generated queries and $74\%$ of real-user queries were identified as natural. The Inter Annotator Agreement (IAA), measured using Fleiss-Kappa \cite{cite-key-fleiss}, was poor ($0.04$) when the users evaluated model-generated sentences. In comparison, when they evaluated queries entered by real users, IAA was better ($0.34$) between the three annotators' ratings and it ranged from fair ($0.22$) to moderate ($0.52$) agreement between each pair of annotators. Higher IAA and higher percentage of queries identified as ``natural'' imply that real-user queries are more natural and distinguishable than queries sampled from pre-trained $Seq2Seq_{NMT}$ model.

\section{Experiment Setup and Results}
\label{sec:experiment-setup-results}
This section describes the experimental setup to train and evaluate the naturalness-estimator, supervised and DRL query generation models.

\subsection{Data}
\label{sec:data}
From user search-query logs, we randomly sampled 0.61 million (90\% train), 34k (5\% valid) and 34k (5\% test) query pairs to train the supervised $Seq2Seq_{NMT}$ and DRL models. Dataset size to train the naturalness-estimator model is 5x the aforementioned amount (See Section \ref{sec:naturalness}). Max-length of a query is 8 and mean-length is \textasciitilde{2} words. Vocabulary size is 32k and out of vocabulary words in validation and test sets are replaced with “<UNK>” unknown-token.

\subsection{Experimental Setup}
\label{sec:experiment-setup}
We implemented all models in Tensorflow \cite{tensorflow2015-whitepaper} and tuned the parameters using Ray[Tune] \cite{liaw2018tune} on Kubernetes \cite{kubernetes} distributed cluster. As described in Section \ref{sec:approach}, the query suggestion policy is initialized with fine-tuned $Seq2Seq_{NMT}$ model. $Seq2Seq_{NMT}$ model parameters are updated using Adam \cite{Kingma2015AdamAM} optimizer and categorical-cross-entropy loss is minimized during training. During inference, six\footnote{In production environment six related queries are suggested for each user query.} queries are generated per input query ($q_i$) using beam-search \cite{DBLP:journals/corr/FreitagA17} decoding. Negative examples to train the two-layered BiLSTM contextual-naturalness-estimator are obtained from pre-trained $Seq2Seq_{NMT}$ model. At inference, naturalness probability ($D_\phi(y^{bk}_{1:T}$)) is obtained from the output of fully-connected layer with last time-step's hidden state as its input.

The initial policy is fine-tuned using REINFORCE policy-gradient algorithm, using future-reward described in Section \ref{sec:reward-formulation}. During training, ``K'' samples for MC roll-out are generated using beam-search or from categorical distribution of inferred word probabilities at each time-step (See Figure \ref{fig:model}).  DRL model training stability is monitored using reward weighted Negative Log Likelihood convergence performance, with $\frac{1}{KB} \Sigma_{(k \epsilon K, b \epsilon B)} [- R_{D_\phi}(y^{bk}_{1:T})*log (G_{\theta}(y^{bk}_{1:T}))]$ as the computed loss at each model training step. 
%\begin{equation}
%\label{eq:nll}
%\frac{1}{KB} \Sigma_{(k \epsilon K, b \epsilon B)} [- R_{D_\phi}(y^{bk}_{1:T})*log (G_{\theta}(y^{bk}_{1:T}))]
%\scriptsize
%\small
%\end{equation}

We use SGD optimizer \cite{10.2307/2236690} to update the weights of the agent (Equation \ref{eq:sgd}). Appendix Figure \ref{fig:nll} shows the convergence performance of the DRL model for different values of unnaturalness penalty ($\eta$), number of MC samples ($K$) and choice of sampling strategy.  Complete set of hyper-parameters we tuned are in Appendix Table \ref{tab:hyper-parameter}. Best combination of hyper-parameters are chosen are based on performance on validation set (See Appendix Table \ref{tab:optimal-hyper-parameter}). 

\subsection{Evaluation Metrics}
\label{sec:metrics}
The binary ``natural/unnatural'' class prediction performance of the contextual naturalness estimator is evaluated using \textit{F1}\footnote{F1 = $\frac{2*Precision*Recall}{Precision + Recall}$} score and  \textit{Accuracy}\footnote{Categorical accuracy: calculates how often predictions match one-hot labels.} metrics. We use the mean of the following metrics calculated on the test set, to evaluate the relevance, engagement, accuracy and diversity of generated queries.

\begin{itemize}[leftmargin=*]
\item \textbf{Sessions with positive user-action ($Sessions^{+}@6$)}:  Long-term binary engagement metric indicating if recommended queries lead to a successful session. Its value is ``1'',  if any of the six generated queries belong to a search-session in test-data with an associated down-stream positive user action (Section \ref{sec:reward-formulation}).
\item \textbf{Unique@6}: Diversity metric indicating the percentage of unique sentences in (six) query suggestions made per query $q^{test}_{i}$. Queries containing unknown word token (``<UNK>'') are filtered out as only high-quality suggestions are presented to the end user.
\item \textbf{Precision@6}: Measures relevance with respect to the query a user would enter next. Is ``1'' if ($q^{test}_{i+1}$) is in the set of six query suggestions made for ($q^{test}_i$) and ``0'' otherwise.
\item \textbf{Word-repetitions per sentence ($Repetitions_S$)}: Fraction of word repetitions per generated query ($S$). Unwanted word repetitions lead to lower quality.
\item \textbf{Prior Sentence Probability ($P_S$)}: $P_S = \Sigma_{w_i \epsilon S} log(p_{w_i})$, measures the prior sentence probability. $p_{w_i}$ is prior word probability defined in Section \ref{sec:naturalness}. Lower sentence probability indicates higher diversity as generated queries contain less frequent words.
\end{itemize}

\subsection{Results}
\label{sec:result}
The contextual-naturalness-estimator achieved $90\%$ accuracy and $80\%$ F1 performance on test set. Table \ref{tab:results} shows the performance of supervised ($Seq2Seq_{NMT}$) and proposed DRL model on the five metrics mentioned in previous section. $DRL_{beam}$ and $DRL_{sampling}$ use ``beam-search'' and ``sampling from categorical distribution'' MC sampling strategies respectively (see Section \ref{sec:experiment-setup}). In order assess the impact of applying heuristics to filter and improve quality of suggestions provided by supervised models, we analyzed the performance of $Seq2Seq^-_{NMT}$, which is $Seq2Seq_{NMT}$ model with post-processing filters to remove suggestions with repeated words.
\begin{table}[h!]
%\captionsetup{font=scriptsize}
\vspace{-0.1cm}
\centering
\resizebox{0.40\textheight}{!}{
  \begin{adjustbox}{max width=\textwidth}
  \bgroup
  \def\arraystretch{1.2}
\begin{tabular}{|l|c|c|c|c|c|}
\hline
Model\textbackslash Metric & $Sessions^{+}@6$ & Unique@6 &  Precision@6 & $Repetitions_S$  & $P_S$ \\
\hline
$Seq2Seq_{NMT}$&  0.1108 $\pm$ 0.002  & 5.8244 $\pm$ 0.0045 & 0.0456 $\pm$ 0.0025 & 2.21\% $\pm$ 0.04\% & -6.4442 $\pm$  0.0149 \\
\hline
$Seq2Seq^-_{NMT}$ &  0.1101 $\pm$ 0.001  & 5.5595 $\pm$ 0.0075 & 0.0456 $\pm$ 0.0025 & \textbf{0.00\%  $\pm$ 0.00\%}$\dagger$ & -6.4875 $\pm$  0.0151$\dagger$  \\
\hline
$DRL_{beam}$ &   \textbf{0.1155 $\pm$ 0.002}$\dagger$   & 5.9606 $\pm$ 0.0023$\dagger$ & \textbf{0.0468 $\pm$ 0.0025} &   1.10\% $\pm$ 0.03\%$\dagger$ & \textbf{-6.4897 $\pm$  0.0140}$\dagger$ \\
\hline
$DRL_{sampling}$ &   0.1149 $\pm$ 0.002$\dagger$  & \textbf{5.9956 $\pm$ 0.0007}$\dagger$  & 0.0467 $\pm$ 0.0024 &  0.40\%$\pm$ 0.02\%$\dagger$ & -6.3932 $\pm$  0.0141 \\
\hline
\end{tabular} 
\egroup
\end{adjustbox}
}
%\vspace{-0.1cm}
\caption{\textmd{Mean performance of supervised $Seq2Seq_{NMT}$ and $DRL$ models across all query pairs in test data. Cells show the mean and 95\% confidence interval calculated using t-distribution. Best mean is in bold. $\dagger$ indicates statistically significant improvement over baseline $Seq2Seq_{NMT}$.}}
\vspace{-0.6cm}
  \label{tab:results}
\end{table}

On offline test data set, in comparison to the baseline $Seq2Seq_{NMT}$ model, $Seq2Seq^-_{NMT}$ removed query suggestions with repeated words completely, however the heuristics-based model performed poorly in-terms of diversity (4.5\% relative drop in mean Unique@6) and average number of successful sessions (0.6\% relative drop in mean $Sessions^{+}@6$). On the other hand, both versions of our proposed DRL models outperformed the baseline model on all metrics. DRL variants achieved significant relative improvement in-terms of user-engagement (mean $Session^+@6$) up to 4.2\% ($0.1108 \rightarrow 0.1155$), query suggestions' diversity (mean Unique@6) up to 3\% ($5.8244 \rightarrow 5.9956$), sentence-level diversity (mean Prior Sentence Probability) up to 0.7\% ($-6.4442 \rightarrow -6.4897$) and reduction in errors per sentence up to  82\% ($2.21 \rightarrow 0.40$). Non significant improvement in relevance (mean Precision@6) is not surprising as the supervised $Seq2Seq_{NMT}$ model is also trained with consecutive query pairs.

\section{Conclusions}
\label{sec:conclusion}
In this paper, we proposed a Deep Reinforcement Learning (DRL) framework to improve the quality of related-search query suggestions. Using long-term user-feedback, syntactic relatedness and estimated unnaturalness penalty as reward signals, we fine-tuned the supervised text-generation policy at scale with REINFORCE policy-gradient algorithm. We showed significant improvement in recommendation diversity ($3\%$),  query correctness ($82\%$),  user-engagement ($4.2\%$) over industry-baselines. For future work, we plan to include semantic relatedness as reward. Since the proposed DRL framework is agnostic to the choice of an encoder-decoder architecture, we plan to fine-tune different state-of-the-art language models using our proposed DRL framework.
%\pagebreak
\begin{acks}
Thanks to Cong Gu, Ankit Goyal and LinkedIn Big Data team for their help in setting up DRL experiments on Kubernetes. Thanks to Souvik Ghosh and RL Foundations team for your valuable feedback.
%To Robert, for the bagels and explaining CMYK and color spaces.
\end{acks}
\bibliographystyle{ACM-Reference-Format}
\bibliography{sample-sigconf.bib}
\newpage
\appendix
\onecolumn
\section{Appendices}
\label{sec:appendix}
\vspace{-0.3cm}
\begin{figure}[h]
\captionsetup{font=scriptsize, width=13cm}
\includegraphics[width=0.60\textwidth]{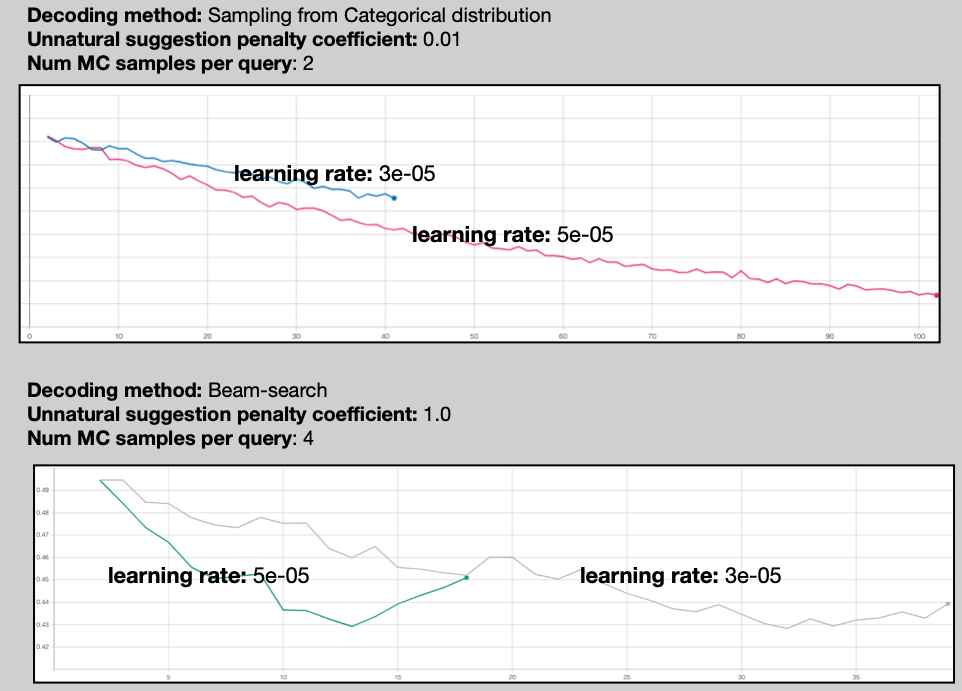}
\vspace{-0.3cm}
\caption{\textmd{Reward weighted Negative log-likelihood convergence performance w.r.t the training epochs of $DRL_{sampling}$ (top figure) and $DRL_{beam}$ (bottom figure) models, for different values of learning-rate. Number of Monte Carlo (MC) samples generated per query during training and unnatural suggestion penalty are from best parameter combination for each MC sampling-method (see Appendix Table \ref{tab:optimal-hyper-parameter}).}}
\label{fig:nll}
\end{figure}

\begin{table*}[h!]
\captionsetup{font=scriptsize, width=13cm}
\vspace{-0.3cm}
%\small
\centering
    \resizebox{0.60\textheight}{!}{
      \bgroup
  \begin{adjustbox}{max width=\textwidth}
  \def\arraystretch{0.7}
  \begin{tabularx}{\textwidth}{X X}
    \toprule
    \multicolumn{1}{l}{\textbf{Model}}  & \multicolumn{1}{l}{\textbf{Hyper parameter and their corresponding ranges}}  \\
    \midrule
$Seq2Seq_{NMT}$& \pbox{18cm}{Batch size:\,$[64, 128, 256, 512]$,\\ \,Sequence Length\,$[4, 5, 6, 7, 8]$, \\ \,Dropout:\,$[0.0-0.4]$,  \\ \,Num RNN Layers:\,$[1, 2, 3]$, \\ \,Hidden-vector length:\, $[128, 256]$} \\\\
Contextual Naturalness Estimator& \pbox{18cm}{Dropout:\,$[0.0 - 0.4]$} \\\\
   % max-depth:\,$4$,\,min-samples-leaf:\,$6$,\,min-samples-split:\,$6$ \\
Deep Reinforcement Learning (DRL) Model & \pbox{18cm}{Learning rate:\,$[1e-04, 1e-06, 1e-05, 2e-05, 3e-05, 4e-05, 5e-05]$,\\ \,Number of samples per input query:\,$[1 - 5]$,  \\ \, Naturalness Penalty Coefficient\,$[1.0, 0.1, 0.01, 0.001]$,  \\ \ Sampling startegy: [Beam Search, Sampling from categorical distribution]}  \\\\
  \bottomrule
  \end{tabularx}
  \end{adjustbox}
  \egroup
  }
  \caption{\textmd{Hyper parameter value ranges we used for training query generation and contextual-naturalness-estimation models. Once hyper-params of Supervised-$Seq2Seq_{NMT}$ are fine-tuned, same model architecture is used for training supervised natural-estimator (encoder) model and DRL (encoder-decoder) agent. Hence batch-size, sequence length, number of rnn layers and hidden vector length remain consistent across all three models.}}
  \label{tab:hyper-parameter}
\end{table*}

\begin{table*}[h!]
\vspace{-0.7cm}
\captionsetup{font=scriptsize, width=13cm}
%\small
\centering
    \resizebox{0.60\textheight}{!}{
      \bgroup
  \begin{adjustbox}{max width=\textwidth}
  \def\arraystretch{0.7}
  \begin{tabularx}{\textwidth}{X X X}
    \toprule
    \multicolumn{1}{l}{\textbf{Model}}  & \multicolumn{1}{l}{\textbf{Best set of hyper-parameters}}  & \multicolumn{1}{l}{\textbf{Criteria}}  \\
    \midrule
$Seq2Seq_{NMT}$ & \pbox{18cm}{Batch size:\,$256$,\\ \,Sequence Length\,$7$, \\ \,Dropout:\,$0.2$,  \\ \,Num RNN Layers:\,$2$, \\ \,Hidden-vector length:\, $256$} & Min categorical cross entropy loss\\\\
Contextual Naturalness Estimator & \pbox{18cm}{Dropout:\,$0.0$} & Max F1 on validation set \\\\
$DRL_{beam}$ & \pbox{18cm}{Learning rate:\,$3e-05$,\\ \,Number of samples per input query:\,$4$,  \\ \, Naturalness Penalty Coefficient\,$1.0$} & Negative Log Likelihood convergence criteria to terminate training. Max $Session^+@6$ performance on validation set to pick the best parameter combination (optimal model). \\\\
$DRL_{sampling}$ & \pbox{18cm}{Learning rate:\,$5e-05$,\\ \,Number of samples per input query:\,$2$,  \\ \, Naturalness Penalty Coefficient\,$0.01$} & Negative Log Likelihood convergence criteria to terminate training. Max $Session^+@6$ performance on validation set to pick the best parameter combination (optimal model). \\\\
  \bottomrule
  \end{tabularx}
  \end{adjustbox}
  \egroup
  }
   \caption{\textmd{Optimal hyper-parameter values for query generation and naturalness estimation models with criteria applied on validation-set performance to choose them.}}
     \label{tab:optimal-hyper-parameter}
\end{table*}

%%
%% The acknowledgments section is defined using the "acks" environment
%% (and NOT an unnumbered section). This ensures the proper
%% identification of the section in the article metadata, and the
%% consistent spelling of the heading.

%%
%% The next two lines define the bibliography style to be used, and
%% the bibliography file.
% \bibliographystyle{ACM-Reference-Format}
% \bibliography{sample-base}

%%
%% If your work has an appendix, this is the place to put it.

\end{document}